\documentclass[twocolumn,reprint,superscriptaddress,amsmath,amssymb,aps,pra]{revtex4-2}
\usepackage{units}
\usepackage{graphicx,subfigure}
\usepackage{mathrsfs}
\usepackage{bm}
\usepackage{textcomp}
\usepackage{url}
\usepackage{dsfont}
\usepackage{braket}
\usepackage[compact]{titlesec}
\usepackage[colorlinks=true,linkcolor=magenta,citecolor=blue]{hyperref}%
\usepackage{todonotes}
\usepackage{empheq}
\usepackage{multirow}
\usepackage{titlesec}
\definecolor{mypink1}{rgb}{0.858, 0.188, 0.478}

\usepackage{cancel}
\titleformat{\subsubsection}
 {\normalfont\normalsize\itshape}{\thesubsubsection}{1em}{}
\titlespacing*{\subsubsection}{0pt}{3.25ex plus 1ex minus.2ex}{0ex plus.2ex}
\begin{document}
\title{Tunable optical amplification and group delay in cavity magnomechanics}
\author{Abdul Wahab} 
\email{abdulwahab@mail.ustc.edu.cn}
\affiliation{Department of Physics, Jiangsu University, Zhenjiang, 212013, China}
\author{Muqaddar Abbas}
\affiliation{Ministry of Education Key Laboratory for Nonequilibrium Synthesis and Modulation of Condensed Matter, Shaanxi Province Key Laboratory of Quantum Information and Quantum Optoelectronic Devices, School of Physics, Xi’an Jiaotong University, Xi’an 710049, China}
\author{Xiaosen Yang} 
\email{yangxs@ujs.edu.cn}
\author{Yuanping Chen}
\email{chenyp@ujs.edu.cn}
\affiliation{Department of Physics, Jiangsu University, Zhenjiang, 212013, China}
\date{\today}
\begin{abstract}
In this work, we theoretically investigate the controllable output probe transmission and group delay in a hybrid cavity magnomechanics (CMM) system. The setup comprises a gain (active) cavity and a passive (loss) cavity, which incorporates an optical parametric amplifier (OPA) and two yttrium iron garnet spheres to facilitate magnon-photon coupling. Unlike the single transparency window typically resulting from magnon-photon interactions, we also observe magnomechanically induced transparency due to nonlinear magnon-phonon interactions. Additionally, two absorption dips on either side of the central absorption dip can be asymmetrically modulated into amplification and absorption by varying different system parameters. A $\mathcal{PT}$-symmetric to broken-$\mathcal{PT}$-symmetric phase transition is observed in both balanced and unbalanced gain-to-loss scenarios. Notably, replacing the second passive cavity with an active one mitigates high absorption and introduces effective gain into the system. Our findings reveal that the group delay of the probe light can be adjusted between positive and negative values by modifying various system parameters. This study provides a robust platform for controlling light propagation in CMM systems, highlighting potential applications in optical communication and signal processing.
\end{abstract}
\maketitle
\section{Introduction}\label{sec:Introduction}
Cavity magnomechanics (CMM)~\cite{ZARERAMESHTI20221, 10.1063/5.0083825, PhysRevB.106.054419} has become a rapidly growing field due to its applications in modern quantum technologies~\cite{YUAN20221,10.1063/5.0152543, PhysRevB.109.L041301, Zuo_2024}. In particular, a new type of hybrid system, primarily relying on collective spin excitations of yttrium iron garnet (YIG) has garnered significant attention and established a platform for quantum information science owing to its robust spin-spin exchange interactions, high spin density, as well as low dissipation rates~\cite{Lachance-Quirion_2019, ZHANG2023100044}. Benefiting from the dynamics of YIG results in promising applications, such as spin current manipulation~\cite{Chumak2015, PhysRevLett.118.217201}, long-time memory ~\cite{Zhang2015, PhysRevLett.127.183202}, microwave-to-optical conversion~\cite{PhysRevB.102.064418, Chai:22}, manipulation of magnons and photons via exceptional points~\cite{PhysRevLett.123.237202, PhysRevLett.124.053602, PhysRevLett.125.147202}, magnon-induced nonreciprocity~\cite{PhysRevLett.123.127202, PhysRevApplied.14.014035}, enhanced tripartite interactions~\cite{PhysRevLett.130.073602}, quantum entanglement~\cite{PhysRevResearch.1.023021,10.1063/5.0015195}, and precision measurements~\cite{PhysRevLett.125.117701,doi:10.1126/science.aaz9236}, etc.

One of the fundamental principles in quantum mechanics is the Hermiticity of physical observables, particularly the Hamiltonian operator. This principle ensures that the eigenvalues are real and that probability is conserved. Interestingly, a broad range of non-Hermitian Hamiltonians can exhibit completely real spectra~\cite{El-Ganainy2018,Ji_2024}. This is especially true for Hamiltonians in parity-time~($\mathcal{PT}$)-symmetric systems.~$\mathcal{PT}$-symmetric systems with gain/loss have received considerable interest in the field of quantum optics~\cite{doi:10.1126/science.aar7709,PhysRevLett.126.230402}, and shown great potential in optomechanical systems~\cite{RevModPhys.88.035002,Longhi_2017}. Remarkably, such systems possess a phase transition~\cite{PhysRevLett.125.183601} from the $\mathcal{PT}$-symmetric phase to the broken-$\mathcal{PT}$-symmetric phase, which will take place at the exceptional point (EP), where the eigenvalues and the associated eigenvectors coalesce and some intriguing physical phenomena may arise. Optical and quantum optical systems are natural test beds for $\mathcal{PT}$-symmetric systems~\cite{PhysRevLett.100.103904,PhysRevLett.103.093902}, which has resulted in showing several exotic phenomena, such as unidirectional invisibility~\cite{Longhi_2011,Feng_2013}, nonreciprocal light propagation~\cite{Peng_2014}, optical light stopping, and sensing~\cite{PhysRevLett.120.013901}. 

New attempts have been made to investigate non-Hermitian physics in magnomechanic quantum systems. The second-order exceptional point is found in a two-mode cavity-magnonic system in which the cavity mode's gain is produced by coherent perfect absorption~\cite{Zhang2017}. Experiments have been carried out to investigate the engineering of anti-$\mathcal{PT}$-symmetry, focusing on the adiabatic elimination of the cavity field and the dissipative coupling between two magnon modes~\cite{PhysRevApplied.13.014053}. In addition to their distinctive spectral responses, these $\mathcal{PT}$-symmetric systems can control the transmission of the output microwave field~\cite{Wang:23,PhysRevApplied.12.034001}. The essential technique underlying such an application ismagnomechanically induced transparency (MMIT)~\cite{Bayati:24,PhysRevA.102.033721}, where a narrow spectral hole appears inside the absorption spectrum due to significant magnon-photon interaction. Further research has shown that weak magnon-phonon coupling can result in multiple transmission windows by altering the phase and amplitude of the applied magnetic field~\cite{10.1063/5.0028395}. Meanwhile, cavity magnonic system was also utilized to investigate novel phenomena and applications~\cite{Sarma_2021,PhysRevLett.127.087203}, such as, nonreciprocal double-carrier frequency combs~\cite{WANG2023114137}, bistability or multistability of magnon–polariton~\cite{PhysRevLett.129.123601}, magnon-induced chaos~\cite{PhysRevE.101.012205}, and so on.

Optomechanically induced transparency (OMIT) is well-known for its application in investigating the slowing and storing of light within a cavity~\cite{PhysRevA.107.033507,PhysRevLett.107.133601}. Leveraging~$\mathcal{PT}$-symmetry in optomechanical systems improves control over light transmission~\cite{jing2015optomechanically,Li2016}, and enables for light conversion from subluminal to superluminal. However, their work may encounter experimental difficulties because the gain in the adjunct cavity could cause instability in the entire system~\cite{Vanderhaegen2021}. The CMM system has significant advantages over optomechanical systems, including strong hybridization between the magnon and photon modes, and a substantially longer lifetime for the phonon mode compared to cavity optomechanics systems. Enhancing the phonon lifetime allows for the transmission of wider spatiotemporal pulses with minimal distortion, enabling them to be stored for longer periods as mechanical excitations. Additionally, CMM systems provide enhanced tunability, as the magnon frequency can be adjusted by varying the external magnetic field. Leveraging these advantages, there has recently been increasing interest in achieving gain in various components of a CMM system~\cite{PhysRevA.99.043803}. This provides a framework for developing a ~$\mathcal{PT}$-symmetry CMM system~\cite{Zhang2017}.

Furthermore, it has been demonstrated that the combination of nonlinear optics and CMM systems has led to numerous physical phenomena to strengthen quantum effects~\cite{Liu:22, PhysRevA.105.063704}. Inserting an optical parametric amplifier (OPA) into the cavity causes an optical amplification~\cite{PhysRevA.100.043824,WAHAB2024115436}, enhancement of optical nonlinearity~\cite{Shahidani:14}, and normal mode splitting~\cite{PhysRevA.80.033807}. Therefore, the interaction between OPA and the cavity can not only alter the dynamic instabilities~\cite{PhysRevA.93.033835} but also find applications such as cooling enhancement~\cite{PhysRevA.79.013821}, and magnon blockade~\cite{PhysRevA.109.043712}. Furthermore, by altering the group delay of the output probe light, one may obtain the slow/fast light effects~\cite{Kong:19,Wahab2024}.

\begin{figure*}
\centering
\includegraphics[width=0.90\linewidth]{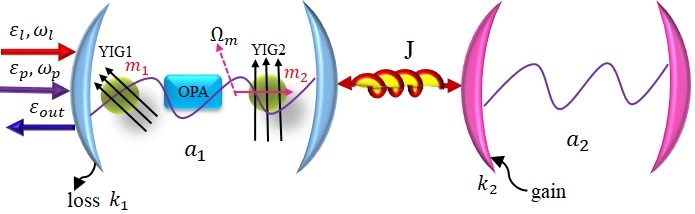}
\caption{Schematic configuration of hybrid nonlinear CMM system. The system is made up of two connected microwave cavities. A  passive (loss) cavity ($a_1$) that contains an OPA and two highly polished single-crystal ferromagnetic YIG spheres ($m_{1}, m_{2}$), and an active (gain) cavity ($a_2$). The $\kappa_1$ and $\kappa_2$ represent the loss and gain rates, respectively. Both the cavities can be interconnected through the optical tunneling rate $J$. The loss cavity is driven by a strong external pump and a weak probe laser field. The magnon $m_{2}$ is directly controlled by a microwave control field with frequency~$\Omega_{m}$. Detailed information about the schematic configuration can be shown in ~\S\ref{section: Model}.} 
\label{model}
\end{figure*}

In this study, leveraging the MMIT effect, we systematically investigate the tunable optical transmission and its associated group delay in a CMM system. The system consists of two optically coupled microcavities: a gain-assisted auxiliary cavity and a passive cavity, which includes an OPA and two YIG spheres driven coherently by a bichromatic laser field. By applying a perturbation approach to solve the Heisenberg-Langevin equations, we derive an analytical solution for the optical transmission. Our calculations demonstrate that, in addition to the single transparency window typically caused by magnon-photon interactions, we also observe MMIT due to nonlinear magnon-phonon interactions. Furthermore, we find that two additional absorption dips, located on either side of the central absorption dip, can be asymmetrically modulated into amplification and absorption by varying different system parameters. We establish that the transition from a broken $\mathcal{PT}$-symmetry phase to an unbroken $\mathcal{PT}$-symmetry phase can be achieved through effective electromechanical coupling, considering both cavity loss and gain. Notably, the transmission rate is significantly enhanced in the active-passive configuration compared to the passive-passive scenario, indicating that the gain from the auxiliary cavity contributes to increasing the transmission of the probe light. Finally, by numerically computing the group delay of the probe field, we show that it is straightforward to achieve the transition between slow and fast light. These attributes suggest that CMM devices can serve as powerful tools for controlling light propagation, with potential applications in optical communication.~\cite{PhysRevLett.92.253201}, and long-lifetime quantum memories~\cite{PhysRevLett.107.133601}.

This work is organized as follows: In~\S\ref{section: Model} we present our proposed theoretical model and derive the steady-state solutions of our system. The results of our research are provided in~\S\ref{section:Results}. Finally, in~\S\ref{section:Conclusions} 
we provide a comprehensive conclusion that encapsulates our research findings.
To ensure the completeness of our work, we provide detailed calculations of algebraic equations in~\S\ref{Appendex}. 
\section{Proposed theoretical model}\label{section: Model}
This section discusses the dynamics of our proposed theoretical model, depicted in Fig.~\ref{model}. Our model comprises two coupled microwave cavities ($a_1$ and $a_2$). Cavity~$a_1$ is treated as a passive (loss) cavity, with the loss rate $\kappa_1$ containing a degenerate OPA and two highly polished single-crystal ferromagnetic YIG spheres (with a 250~$\mu$m diameter as in Ref~\cite{doi:10.1126/sciadv.1501286}). We refer to this cavity as a CMM resonator. Applying an external uniform magnetic field to each YIG sphere activates the magnon mode, which interacts with the microwave cavity mode through magnetic dipole interaction~\cite{PhysRevLett.116.223601, PhysRevLett.123.107702, PhysRevLett.125.237201}. The external bias magnetic field causes deformation of the lattice structure of the YIG spheres, resulting in magnon-phonon interaction, and vice versa~\cite{PhysRev.110.836}. The strength of the single-magnon magnomechanical interaction is exceedingly small~\cite{doi:10.1126/sciadv.1501286} and it is influenced by both the direction of the applied bias field and the diameter of the sphere. 
It is feasible to disregard the magnomechanical interaction of the YIG1 sphere by either employing a bigger sphere or changing the direction of the applied biased magnetic field~\cite{Li_2019}. Here, we suppose that the bias field on YIG1 is such that the single-magnon magnomechanical interaction is weak enough to be ignored~\cite{doi:10.1126/sciadv.1501286}. The active (gain) cavity~($a_2$) with the gain rate $\kappa_2$ is coupled to the passive (loss) cavity via optical tunneling rate $J$. The CMM resonator is driven by two external optical laser fields: a weak probe and a strong pump field. The symbols $\varepsilon_p$ and $\varepsilon_l$ denote the amplitudes, $\omega_p$ and $\omega_l$ represent the frequencies and $\phi_p$, $\phi_l$ correspond to the phases of the probe and pump fields, respectively. A weak external microwave field with frequency $\Omega_{m}$ and the phase $\phi_m$ is directly drives the YIG2 sphere to enhance the magnon-phonon coupling. 

In this study, we assume that the diameters of the YIG spheres are significantly smaller than the wavelengths of the applied fields, allowing us to safely neglect the influence of light radiation pressure in our system~\cite{PhysRevLett.121.203601,PhysRevA.99.021801}. We also ignored the nonlinear Kerr term $K\hat{m}_{j}^{\dagger}\hat{m}_{j}^{\dagger}\hat{m}_{j}\hat{m}_{j}$ which may occur due to a strongly driven magnon mode~\cite{PhysRevLett.120.057202,PhysRevB.94.224410}. The total Hamiltonian of the system reads:
\begin{equation}
\begin{aligned}
\hat{H}_{\text {t}}= & \hbar \omega_1 \hat{a}_1^{\dagger} \hat{a}_1+\hbar \omega_2 \hat{a}_2^{\dagger} \hat{a}_2+\hbar \omega_b \hat{b}^{\dagger} \hat{b}+\hbar J\left(\hat{a}_1^{\dagger} \hat{a}_2+\hat{a}_2^{\dagger} \hat{a}_1\right)\\&
+\hbar\sum_{j=1}^2\left[\omega_{m j} \hat{m}_j^{\dagger} \hat{m}_j+g_j\left(\hat{a}_1+ \hat{a}_1^{\dagger}\right)\left(\hat {m}_j+\hat{m}_j^{\dagger}
\right)\right]
\\& +\hbar g_{m b} \hat{m}_{2}^{\dagger} \hat{m}_{2}\left(\hat{b}^{\dagger}+\hat{b}\right)+i \hbar G\left(e^{i \theta} \hat{a}_1^{\dagger 2}e^{-2i \omega_{l}t}-\text { H.c.}\right)\\&+i \hbar \sqrt{ 2\eta_a \kappa_1} \varepsilon_l\left(\hat{a}_1^{\dagger}e^{-i \omega_l t-i \phi_l}-\text { H.c.}\right)\\&+ i \hbar \left(\varepsilon_m {m}_{2}^{\dagger} e^{-i \Omega_m t-i \phi_m}-\text{H.c.}\right)\\&+i \hbar \sqrt{2 \eta_a \kappa_1} \varepsilon_p\left(\hat{a}_1^{\dagger} e^{-i \omega_p t-i \phi_p}-\text { H.c.}\right).\label{hamiltonia-1a}
\end{aligned}
\end{equation}
We assume that both cavities in our CMM system have the same resonance frequency ($\omega_{1} = \omega_{2} = \omega_{a}$)~\cite{LIAO2023112978}. Applying the rotating-wave approximation to the system, the interaction terms simplify as $g_j(\hat{a}_1+ \hat{a}_1^{\dagger})(\hat {m}_j+\hat{m}_j^{\dagger}) \rightarrow g_j(\hat{m}_j^{\dagger} \hat{a}_1 + \hat{m}_j \hat{a}_1^{\dagger})$, by neglecting the fast  oscillating terms $g_j(\hat{a}_1 \hat{m}j + \hat{a}1^{\dagger} \hat{m}j^{\dagger})$. This is valid when~$\omega_{a}\gg g_j,\kappa_1, \kappa_{m j}$ which is the case to be considered in the present work ($\kappa_1$ and $\kappa_{m j}$ are the decay rates of the cavity and magnon modes, respectively). To proceed, it is convenient to shift to a frame rotating at the drive frequency~$\omega_{l}$. Following the transformation $\hat{H}_{\text {rot}}=R\hat{H}_{\text {t}}R^{\dagger}+i \hbar (\frac{\partial R}{\partial t})R^{\dagger}$ with $R=e^{i \omega_l(\hat{a}_1^{\dagger} \hat{a}_1+\hat{a}_2^{\dagger} \hat{a}_2+\hat{m}_j^{\dagger}\hat{m}_j)t}$, the Hamiltonian in Eq.~(\ref{hamiltonia-1a}) can be rewritten as~\cite{PhysRevX.11.031053, PhysRevA.103.053501}:
\begin{equation}
\begin{aligned}
\hat{H}_{\text {rot}}= & \hbar \Delta_a\left(\hat{a}_1^{\dagger} \hat{a}_1+\hat{a}_2^{\dagger} \hat{a}_2\right)+\hbar \omega_b \hat{b}^{\dagger} \hat{b}+\hbar J\left(\hat{a}_1^{\dagger} \hat{a}_2+\hat{a}_2^{\dagger} \hat{a}_1\right)\\&
+\hbar\sum_{j=1}^2\left[\Delta_{m j} \hat{m}_j^{\dagger} \hat{m}_j+g_j\left(\hat{m}_j^{\dagger} \hat{a}_1+m_j \hat{a}_1^{\dagger}\right)\right]
\\& +\hbar g_{m b} \hat{m}_{2}^{\dagger} \hat{m}_{2}\left(\hat{b}^{\dagger}+\hat{b}\right)+i \hbar G\left(e^{i \theta} \hat{a}_1^{\dagger 2}-e^{-i \theta} \hat{a}_1^2\right)\\&+i \hbar \sqrt{ 2\eta_a \kappa_1} \varepsilon_l\left(\hat{a}_1^{\dagger}-\hat{a}_1\right)+ i \hbar \left(\varepsilon_m e^{-i \delta_{ml} t-i \phi_m}\hat{m}^{\dagger}-\text{H.c.}\right)\\&+i \hbar \sqrt{2 \eta_a \kappa_1} \varepsilon_p\left(\hat{a}_1^{\dagger} e^{-i \delta_{pl}t-i \phi_{pl}}-\text { H.c.}\right),\label{hamiltonian}
\end{aligned}
\end{equation}
where $\Delta_{a}:=\omega_{a}-\omega_{l}$ ($\sum_{j=1}^2\Delta_{m j}:=\omega_{m j}-\omega_{l}$) shows the frequency detuning between cavity (magnon) and control field. $\delta_{pl}:=\omega_{p}-\omega_{l}$ stands for the frequency detuning of probe-control field, and  $\delta_{ml}:=\Omega_{m}-\omega_{l}$ shows the detuning of the magnon drive and control fields. $\phi_{pl}=\phi_{p}-\phi_{l}$ 
indicates the relative phase between the probe and pump fields. $\omega_{a}$, $\omega_{b}$, and $\omega_{m}$ are the frequencies that correspond to the resonances of the photon, phonon, and magnon modes, respectively.  The term $\hbar J\left(\hat{a}_1^{\dagger} \hat{a}_2+\hat{a}_2^{\dagger} \hat{a}_1\right)$ indicates the photon exchange coupling strength interaction between the two cavities. $\sum_{j=1}^2 g_{j}$ ($g_{m b}$) signifies the magnon-photon (phonon) coupling rates~\cite{PhysRevB.105.214418}. $G$ represents nonlinear gain, while $\theta$ indicates the phase of OPA, respectively. 
The term $\varepsilon_m = \sqrt{5N/4} B_{0}\gamma$ represents the coupling strength between the magnon and the microwave field; in which~$N= \rho V$ is the total number of spins (with $V$ shows the volume, and ~$\rho= 4.22 \times10^{27}\textup{cm}^{-3}$ is the spin density of YIG sphere)~\cite{PhysRevLett.121.203601,PhysRevLett.124.213604}, $\gamma/2\pi$= 28 GHz/T represents the gyromagnetic ratio, with $B_{0}$ shows the field amplitude. The terms $i \hbar \sqrt{2\eta_a\kappa_1}\varepsilon_l\left(\hat{a}_1^{\dagger}-\hat{a}_1\right)+i \hbar \sqrt{2\eta_a \kappa_1}\left(\hat{a}_1^{\dagger} e^{-i \delta_{pl}t-i \phi_{pl}} \varepsilon_p-\text{ H.c.}\right)$ signify the interaction between the cavity field and two input fields. The terms~$\varepsilon_l =\sqrt{P_{l}/\hbar \omega_l} (\varepsilon_p =\sqrt{P_{p}/\hbar \omega_p})$ represent the amplitude of the strong pump (weak probe) fields, whereas $P_{l}$ and $P_{p}$ are the powers of the corresponding pump and probe fields, respectively. The symbol~$\eta_{a}$ represents the coupling between the CMM resonator and the output port, defined as $\eta_{a}=\kappa_{c_{1}}/\kappa_1$, where $\kappa_{c_{1}}$ is the external decay rate of the cavity and $\kappa_1$ is the total decay rate of the cavity fields, consisting of both intrinsic and external decay rates ($\kappa_1 = \kappa_i + \kappa_{c_{1}}$)~\cite{PhysRevA.110.023502,PhysRevA.92.033823}. In this study, we consider the CMM resonator operating in the critical-coupling regime and set~$\eta_{a}=0.5$ throughout the work~\cite{doi:10.1126/science.1195596}.

\subsection{Quantum dynamics and fluctuations:}
In our study, we primarily focus on the average response of the system; thus by considering the Hamiltonian of Eq.~(\ref{hamiltonian}), and ignoring the input quantum noises~\cite{PhysRevA.86.013815}, the Heisenberg-Langevin equations of our compound CMM system can be expressed as:
\begin{figure*}
\centering
\includegraphics[width=0.33\linewidth]{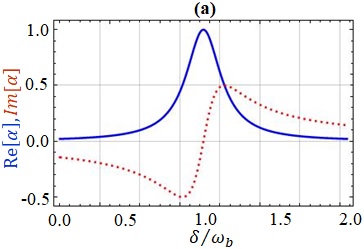}\hfill
\includegraphics[width=0.33\linewidth]{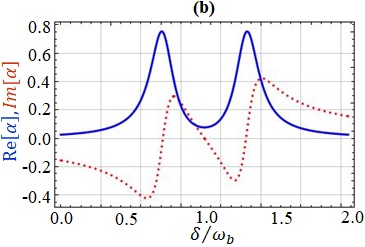}\hfill
\includegraphics[width=0.33\linewidth]{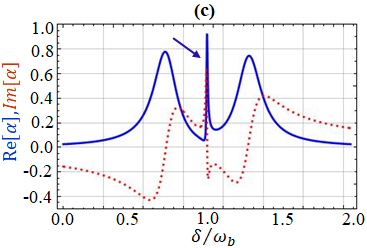}
\caption{~Absorption~Re$[\alpha]$~(blue solid line) and dispersion~$Im[\alpha]$~(red dotted line) profiles of the output probe field are shown versus optical detuning $\delta/\omega_{b}$. We select the coupling strengths as~(a)~$g_{1} =2 \kappa_1$,~$g_{2} =g_{mb}=0$~(b)~$g_{1}=g_{2} =2 \kappa_1$,~$g_{mb}=0$~(c)~$g_{1}=g_{2} =2 \kappa_1$,~$g_{mb}=2\pi$~Hz. The parameter values are considered as: $\omega_c = 2 \pi \times 7.86$~GHz,~$\kappa_1 = \pi \times 3.35 $~MHz,~$\kappa_2 = \kappa_1$,~$D = 250\mu$m,~$\rho= 4.22 \times10^{27}\textup{cm}^{-3}$,~$\hbar = 1.054 \times 10^{-34} \ \mathrm{J \cdot s}$,~$\omega_b = 2 \pi \times 11.42$~MHz,~$\kappa_b = 300 \pi$~Hz,~$\kappa_{m_{1}}= \kappa_{m_{2}}= \pi \times 1.12$~MHz,~$B_{0}=0$,~$N= 3\times10^{16}$,~$\gamma$= $2 \pi \times$ 28 GHz/T,~$P_{l} =10$~mW,~$\varepsilon_p =0.0001 \varepsilon_l$,~$\eta_a=0.5$,~$\Delta_a =\Delta_{m_{1}}= \Delta_{m_{2}}^{\prime} = \omega_b$,~$G = 0$,~$\theta=0$,~$\Phi_m=0$,~and~$J = 0$.}
\label{magnon phonon}
\end{figure*}
\begin{equation}
\begin{aligned}
\left\langle\dot{a}_1\right\rangle= & \left(-i \Delta_a-\kappa_1\right)\left\langle a_1\right\rangle-i \sum_{j=1}^2 g_{j}\langle m_{j}\rangle-i J\left\langle a_2\right\rangle \\
& +\sqrt{ 2\eta_a \kappa_1} \varepsilon_l+\sqrt{2 \eta_a \kappa_1} \varepsilon_p e^{-i \delta_{pl}t-i \phi_{pl}} + 2 G e^{i \theta} a_1^{\dagger}, \\
\langle\dot{m_{1}}\rangle= & \left(-i \Delta_{m 1}-\kappa_{m 1}\right)\langle m_{1}\rangle- ig_{1}\left\langle a_1\right\rangle,
\\
\langle\dot{m_{2}}\rangle= & \left(-i \Delta_{m 2}-\kappa_{m 2}\right)\langle m_{2}\rangle- ig_{2}\left\langle a_1\right\rangle \\
& + \varepsilon_m e^{-i \delta_{ml} t-i \phi_m} -ig_{m b}\langle m_{2}\rangle\left(\left\langle b^{\dagger}\right\rangle+\langle b\rangle\right), \\
\langle\dot{b}\rangle= & \left(-i \omega_b-\kappa_b\right)\langle b\rangle-i g_{m b}\left\langle m_{2}^{\dagger}\right\rangle\langle m_{2}\rangle, \\
\left\langle\dot{a}_2\right\rangle= & \left(-i \Delta_a+\kappa_2\right)\left\langle a_2\right\rangle-i J\left\langle a_1\right\rangle,\label{hamiltonian-1}
\end{aligned}
\end{equation}

where $\sum_{j=1}^2\kappa_{m j}$ and $\kappa_b$ denotes the decay rates of magnons and phonons. Keep in mind that $\kappa_1 > 0$, $\kappa_2 > 0$ in Eq.~(\ref{hamiltonian-1}) is associated to passive-active compound CMM resonator system, while $\kappa_1 > 0$, $\kappa_2 < 0$ is associated to a passive-passive system~\cite{PhysRevA.108.033517}.

Relative to the intensity of the controlled field, we assume that the intensities of the external magnetic driving fields and the weak probe field satisfies the condition $|\varepsilon_{p}|\ll \varepsilon_{l}$ and $|\varepsilon|_{m}\ll \varepsilon_{l}$~\cite{PhysRevA.86.013815,PhysRevA.102.023707}. Thus in this case, we can linearize the set of above dynamical equations of our proposed model by assuming each operator is the sum of its mean value and fluctuation i.e.,~$O= O_s+ O_{+} e^{-i \delta t}+O_{-} e^{i \delta t}$, where $O=a_1, a_2, b, m_{1}, m_{2}$~\cite{PhysRevA.102.033721,PhysRevA.83.043826}, the first term $O_s$ represents the steady-state values and $\delta{O}= O_{+} e^{-i \delta t}+O_{-} e^{i \delta t}$ depicts the small fluctuating terms. By using perturbation expansion ansatz in Eq.~(\ref{hamiltonian-1}) and considering the time derivative equal to zero we can obtain the steady-state solution as:
\begin{subequations}
\begin{equation}
\begin{aligned}
a_{1 s} & =\frac{(\lambda_{2}+ 2 G e^{i \theta}){\sqrt{2\eta_a \kappa_1} \varepsilon_l}}{ \lambda_{2} \lambda_{3}},
\end{aligned}
\end{equation}
\begin{equation}
m_{1s} = \frac{-i g_{1} a_{1 s}}{i \Delta_{m 1} + \kappa_{m 1}}, 
\end{equation}
\begin{equation}
m_{2s} = \frac{-i g_{2} a_{1 s}}{i \Delta_{m 2}^{\prime} + \kappa_{m 2}}, 
\end{equation}
\begin{equation}
\begin{aligned}
b_s  =\frac{-i g_{m b}\left|m_{2s}\right|^2}{i \omega_b+\kappa_b}, 
\end{aligned}
 \end{equation}
\begin{equation}
\begin{aligned}
a_{2 s} =\frac{i J a_{1 s}}{\left(-i \Delta_a+\kappa_2\right)}. \label{hamiltonian-2} 
\end{aligned}
\end{equation}
\end{subequations}

Here we define: $\lambda_{1} = \frac{{(\kappa_1 + i \Delta_a)((\kappa_2 - i \Delta_a)) - J^2}}{{(\kappa_2 - i \Delta_a)}} $,~$\lambda_{2} = (\kappa_1 - i \Delta_a) + (\frac{g_1^2}{-i \Delta_{m 1} + \kappa _{m 1}}) + (\frac{g_2^2}{-i \Delta_{m 2} + \kappa _{m 2}}) - (\frac{J^2}{\kappa_2 + i \Delta a})$,~$\lambda_{3} = (\frac{\lambda_{1} \lambda_{2} - 4 G^2}{\lambda_{2}}) + (\frac{g_1^2}{i \Delta_{m 1} + \kappa _{m 1}}) + (\frac{g_2^2}{i \Delta_{m 2} + \kappa _{m 2}})$. By performing a series of mathematical calculations and neglecting all higher-order terms $(\delta o \delta o)$ the evolution of the perturbation terms of Eq.~(\ref{hamiltonian-1}) can be shown as:

\begin{equation}
\begin{aligned}
\delta \dot{a}_1= & -\left(i \Delta_a+\kappa_1\right) \delta a_1-i J \delta a_2-i \sum_{j=1}^2 g_{j} \delta m_{j} \\
& +\sqrt{2 \eta_a \kappa_1} \varepsilon_p e^{-i \delta_{pl}t-i \phi_{pl}}+2 G e^{i \theta}\delta a_1^{\dagger}, \\
\delta \dot{m_{1}}= & -\left(i \Delta_{m_{1}}+\kappa_{m_{1}}\right) \delta {m_{1}}-i g_{1} \delta a_1, \\
\delta \dot{m_{2}}= & -\left(i \Delta_{m_{2}}^{\prime}+\kappa_{m_{2}}\right) \delta {m_{2}}-i g_{2} \delta a_1-i F \delta b\\
&-i F \delta b^{\dagger}+ \varepsilon_m e^{-i \delta_{ml} t-i \phi_m}, \\
\delta \dot{b}= & -\left(i \omega_b+\kappa_b\right) \delta b-i F \delta m_{2}^{\dagger}-i F^* \delta m_{2}, \\
\delta \dot{a}_2= & -\left(i \Delta_a-\kappa_2\right) \delta a_2-i J \delta a_1, \label{Heisenberg-Langevin-3} 
\end{aligned}
\end{equation}
here, $F =g_{m b} m_{2s}$, and $\Delta_{m_{2}}^{\prime}= \Delta_{m_{2}}+ g_{m b} (b_s + b_s^*)$ signifies the effective strength of the magnon-phonon interaction, and effective magnon detuning respectively.

To determine the amplitudes of the first-order side band, we expand the small fluctuation terms in Eq.~(\ref{Heisenberg-Langevin-3}) by using the following ansatz as~\cite{PhysRevA.107.063714,PhysRevA.86.013815}:
\begin{equation}
\begin{aligned}
\delta {a}_1 =& A_{1+} e^{-i \delta t}+A_{1-} e^{i \delta t} ,\\
\delta {m}_1 =&M_{1+} e^{-i \delta t}+M_{1-} e^{i \delta t}, \\
\delta {m}_2 =&M_{2+} e^{-i \delta t}+M_{2-} e^{i \delta t},\\
\delta b =&B_{1+} e^{-i \delta t}+B_{1-} e^{i \delta t}, \\
\delta a_{2} =&X_{1+} e^{-i \delta t}+X_{1-} e^{i \delta t}. \label{Heisenberg-Langevin-6}
\end{aligned}
\end{equation}
We consider the magnon driving field to
become resonant with the probe field frequency which leads us to consider~$\delta= \delta_{pl} = \delta_{ml}$~\cite{10.1063/5.0028395,Liu_2024}, and $A_{i+}$ and $A_{i-}$ denotes the~$i$th cavity generated probe field and four-wave mixing field amplitudes, respectively.

Now by employing Eq.~(\ref{Heisenberg-Langevin-6}) into Eq.~(\ref{Heisenberg-Langevin-3}) and comparing the coefficients of similar order, one can get the amplitudes of the first  ordered side-bands (detail calculations and the explicit constants are presented in~\S\ref{Appendex}) as~\cite{PhysRevA.107.063714,PhysRevA.86.013815}:
After solving Eqs.~(\ref{Heisenberg-Langevin-7}) we derive the coefficients for the first lower sidebands. This demonstrates the characteristics of our system. Therefore, the coefficient of the $A_{1+}$, which indicates the output probe field amplitude of the CMM resonator can be written as:
\begin{equation}
\begin{aligned}
 A_{1+} =  \left(\frac{\beta_{4}\varepsilon_m e^{i  \Phi}-\sqrt{2 \eta_a \kappa_1} \varepsilon_p}{\alpha_{13}}\right). \label{Heisenberg-Langevin-new}
\end{aligned}
\end{equation}
Here, $\Phi=\phi_m-\phi_{pl}$ corresponds to the relative phase of the applied fields. Furthermore, by employing the input/output relationship of the cavity~$s_{\text {out}}=s_{\text {in}}-\sqrt{2\eta \kappa_1}a_{1}$~\cite{PhysRevA.98.063840}, we get the normalized output probe field intensity from the CMM resonator as: 
\begin{equation}
\begin{aligned}
T = |t_{p}|^{2}= \left|\frac{\sqrt{2\eta_a \kappa_1}A_{1+}}{\varepsilon_p}-1\right|^2.
\end{aligned}
\end{equation}
The real and imaginary components of the $\alpha$ ($\alpha = t_{p}+1$) can be quantified as the absorption and dispersion of the probe field at the probe frequency, respectively.

\begin{figure*}
\centering
\includegraphics[width=0.43\linewidth]{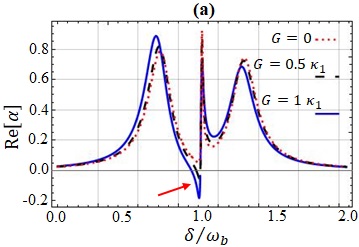}\hspace{2em}
\includegraphics[width=0.43\linewidth]{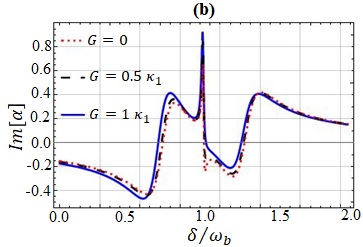}\par\medskip
\caption{(a) Absorption~Re$[\alpha]$ and (b) dispersion~$Im[\alpha]$~profiles of the output probe field are shown versus optical detuning $\delta/\omega_{b}$ for the different selected parameters of OPA gains as $G = 0$,~$G=0.5\kappa_1$,~$G=1\kappa_1$
and~$\theta=0$. All other parameters remains same as in Fig.~\ref{magnon phonon}(c) except for~$B_{0}=5\times10^{-10}$T.}
\label{REIMG}
\end{figure*}
\section{Results}\label{section:Results}
In this part, we provide our numerical findings to investigate the transmission and group delay of the output probe field in a cavity-magnon system by employing a set of experimentally feasible parameter values~\cite{doi:10.1126/sciadv.1501286,PhysRevB.105.214418}, unless stated differently. We use the following values, i.e., the degenerate microwave cavity frequency~$\omega_c = 2 \pi \times 7.86$~GHz, the first cavity decay rate $\kappa_1 = \pi \times 3.35 $~MHz, and the coupling parameter $\eta_a=0.5$. The values for the YIG sphere are taken as: the diameter~$D = 250\mu$m, the spin density~$\rho= 4.22 \times10^{27}\textup{cm}^{-3}$. The frequency of the phonon mode~$\omega_b = 2 \pi \times 11.42$~MHz, and decay rate~$\kappa_b = 300 \pi$~Hz. The magnon decay rate~$\kappa_{m_{1}}= \kappa_{m_{2}}= \pi \times 1.12$~MHz, gyromagnetic ratio~$\gamma$= $2 \pi \times$ 28 GHz/T,
and~$N= 3\times10^{16}$. In addition, we take the power of the pump field~$P_{l} =10$~mW, and the amplitude of the probe filed $\varepsilon_p =0.0001 \varepsilon_l$, with~$\Delta_a =\Delta_{m_{1}}= \Delta_{m_{2}}^{\prime} = \omega_b$.

In~\S\ref{Effect of magnon-photon and magnon-phonon  coupling strength on MMIT window profiles} we study the effect of magnon-photon and magnon-phonon coupling strengths on MMIT window profiles. Next, in~\S\ref{Effect of microwave field and OPA on the transmission and second-order sideband generation} we examine how the amplitude of the microwave field, its relative phase, and the nonlinear parameter OPA impact the transmission rate of the output probe field. Then, in~\S\ref{Controllable phase transition and MMIT in passive-active CMM system} we present our analysis to evaluate the controllable phase transition and MMIT in a passive-active CMM system. Furthermore, in~\S\ref{Optimization of group delays in an active-passive compound CMM system} we represent our results for optimizing group delays in an active-passive compound CMM system. Finally, in~\S\ref{Possible experimental setup} we briefly discuss the possible experimental realization of our proposed scheme.
\subsection{Effect of magnon-photon and magnon-phonon coupling strengths, and OPA on MMIT window profiles}\label{Effect of magnon-photon and magnon-phonon  coupling strength on MMIT window profiles}
In this subsection, we will explore the physics behind the formation of MMIT window profiles by systematically analyzing the influence of magnon-photon and magnon-phonon coupling strengths in our model. Figure~\ref{magnon phonon} displays the absorption (dispersion) spectrum of the probe field versus probe detuning ($\delta/\omega_{b}$) for various coupling strengths. In Fig.~\ref{magnon phonon}(a), we consider the scenario where both the magnon-phonon coupling ($g_{mb}$) and the coupling strength of the magnon mode $m_{2}$ ($g_{2}$) with the cavity are absent. As a result, only magnon mode~$m_{1}$ is coupled with the cavity. Following this scenario, we observe a typical Lorentzian-type absorption peak of the output spectrum (as shown by a blue solid line) in Fig.~\ref{magnon phonon}(a), and the corresponding dispersion curve is indicated by red dotted lines. Interestingly, when the coupling strengths of both magnon modes~$m_{1},m_{2}$~($g_{1},g_{2}$) are present, and only the magnon-phonon coupling~($g_{mb}$) is absent, we observe that the typical Lorentzian-type absorption peak at resonance~$\delta=\omega_{b}$ transforms into a transparency window. Additionally, two symmetrical absorption dips occur at $\delta=0.7\omega_{b}$(left) and  $\delta=1.3\omega_{b}$(right) as shown in Fig.~\ref{magnon phonon}(b). The central peak at~$\delta=\omega_{b}$ represents magnon-induced transparency, while the two sideband absorption peaks indicate magnon-induced absorption~(MIA)~\cite{Kong:19}. 
The physical reason for these peaks is the generation of magnon-polaritons due to the photon-magnon interaction in our system. As a result, our entire system is effectively reduced to a cavity magnonic (CM) system~\cite{PhysRevLett.121.137203,10.1063/1.5126600,PhysRevB.100.094415}. 
\begin{figure*}
\centering
\includegraphics[width=0.34\linewidth]{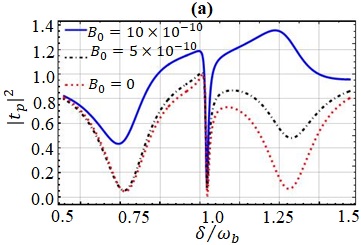}
\includegraphics[width=0.34\linewidth]{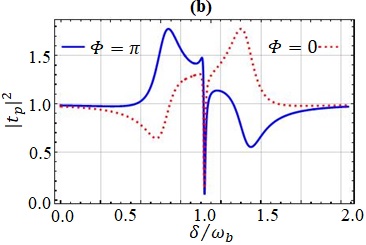}
\includegraphics[width=0.30\linewidth]{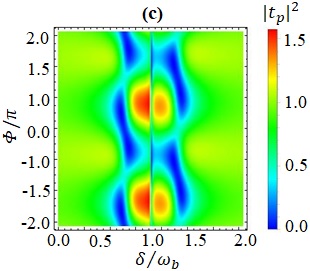}
\caption{~Transmission~$|t_{p}|^{2}$ spectrum of the output probe field versus optical detuning $\delta/\omega_{b}$ is shown for different values of the magnetic field amplitude ($B_{0}$) and the relative phase ($\Phi$) of magnon driving field $\varepsilon_m$.~(a)~The magnetic field amplitude~$B_{0}=0$,~$B_{0}=5\times10^{-10}$T,~and ~$B_{0}=10\times10^{-10}$T. (b)~The relative phase
$\Phi=0$,
$\Phi=\pi$~and~$B_{0}=12\times 10^{-10}$T~(c)~Contour map of transmission rate~$|t_{p}|^{2}$ vs optical detuning $\delta/\omega_{b}$ and relative phase~$\Phi$. All other parameters remains same as in Fig.~\ref{magnon phonon}(c) except for~$B_{0}=5\times10^{-10}$T.}
\label{Transmission a}
\end{figure*}
It is noteworthy that the width of the transparency window and the amplification in our system can be controlled by adjusting the magnon-photon coupling strength~($g_{2}$). However, the results are not shown here. Finally, we consider the scenario where all three couplings are present, i.e., $g_{1}=g_{2}= 2\kappa_{1}$, and $g_{mb}=2\pi$ Hz. In this case, due to the nonzero magnetostrictive interaction, the single MMIT window at the resonance point ($\delta=\omega_{b}$) of the~Fig.~\ref{magnon phonon}(b) splits into double MMIT windows with a central absorption peak, as shown in~Fig.~\ref{magnon phonon}(c). The physical explanation is that the existence of phonon interactions can create additional pathways for destructive interference, leading to additional transparency windows and our entire system reduced to CMM system~\cite{doi:10.1126/sciadv.1501286}.

In Fig.~\ref{REIMG}, we plot the real and imaginary parts of the output probe field as a function of optical detuning $\delta/\omega_{b}$ for different selected OPA gain parameters $G$. It is evident that changes in the parametric gain influence the transparency window. In the absence of OPA ($G=0$), we observe clear absorption in our system. However, introducing OPA~($G\neq0$) converts the positive absorption peak at resonance into a negative peak, indicating gain in our system. The gain provided by the OPA in CMM systems offers substantial benefits, including signal amplification, tunability, sensitivity, and the ability to explore complex physical phenomena. Given these advantages, our work provides greater benefits compared to previously reported studies in~\cite{PhysRevA.102.033721}.
\begin{figure*}
\centering
\includegraphics[width=0.43\linewidth]{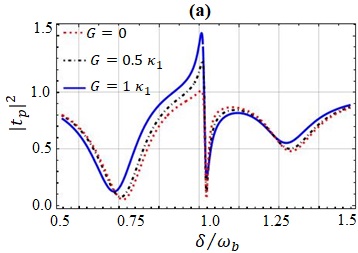}\hspace{2em}
\includegraphics[width=0.43\linewidth]{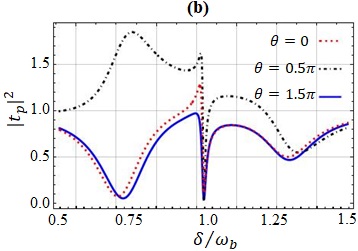}\par\medskip
\caption{~Transmission rate~$|t_{p}|^{2}$ spectrum of the output probe field versus optical detuning $\delta/\omega_{b}$ for different values of OPA gains and phases.~(a)~The OPA gains parameters are chosen as: $G = 0$,~$G=0.5\kappa_1$,~$G=1\kappa_1$,
and~$\theta=0$.~(b)~The phases of OPA are set as:~$\theta=0$, ~$\theta=0.5\pi$,~$\theta=1.5\pi$},~and $G=0.5 \kappa_1$.~All other parameters remains same as in Fig.~\ref{magnon phonon}(c) except for~$\Phi=0$ and~$B_{0}=5\times10^{-10}$T.
\label{Angle}
\end{figure*}
\begin{figure*}
\centering
\includegraphics[width=0.42\linewidth]{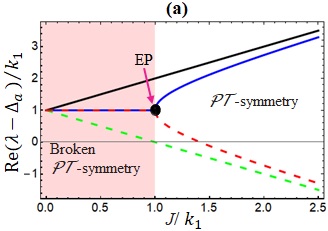}\hspace{2em}
\includegraphics[width=0.42\linewidth]{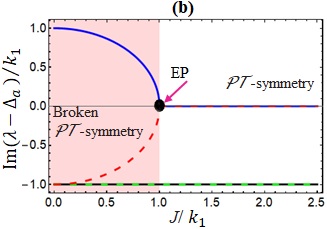}\par\medskip
\includegraphics[width=0.42\linewidth]{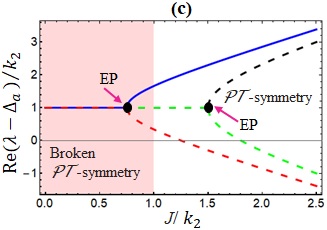}\hspace{2em}
\includegraphics[width=0.42\linewidth]{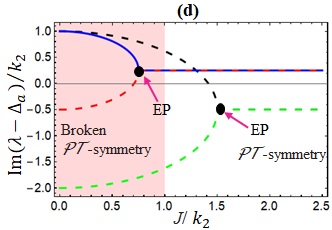}
\caption{~(a)~Real and (b)~Imaginary components of the eigenfrequencies~($\lambda- \Delta_a)/\kappa_1$ are depicted versus coupling strength $J/k_{1}$ for identical gain and loss ($k_{1}=k_{2}$).~(c)~Real and (d)~imaginary components of eigenfrequencies~($\lambda- \Delta_a)/\kappa_2$ are depicted versus coupling strength~$J/k_{2}$ for different gain/loss ratios~$R$, where red dashed and blue solid lines corresponds to~$R = 0.5$, and green (black) dashed lines correspond to~$R = 2$, respectively. Distinct background colors are utilized to differentiate between the unbroken and broken $\mathcal{PT}$-symmetric regimes.}
\label{PT}
\end{figure*}
\subsection{Role of microwave field's amplitude, phase, and OPA on transmission profiles}\label{Effect of microwave field and OPA on the transmission and second-order sideband generation}
In the following, we investigate the impact of the microwave field's amplitude, its relative phase, and the nonlinear parameter OPA on the transmission profiles of the output spectrum. Referring to our discussion about Fig.~\ref{magnon phonon}, the positions of the magnetomechanically induced absorption and the MIA peaks are determined, respectively. Here, we are interested in employing an extra magnon driving field to examine the detailed impacts of the magnon driving parameters on the features of the output field in our CMM system. In~Fig.~\ref{Transmission a}(a), 
We graph the transmission spectrum ($|t_{p}|^{2}$) of the output probe field versus optical detuning ($\delta/\omega_{b}$) for different magnetic field amplitudes ($B_{0}$), keeping the relative phase $\Phi=0$. It is observed that with an increase in the magnetic field amplitude ($B_{0}$), the central absorption peak at the resonance point ($\delta=\omega_{b}$) becomes slightly deeper. There is a slight enhancement in the transparency spectrum on the left side of the resonance point, coupled with a lower absorption dip on the right side of the resonance point ($\delta=\omega_{b}$). Interestingly, when we further increase the magnetic field amplitude ($B_{0}=10\times10^{-10}$T), the symmetric absorption dips on both sides of the resonance point ($\delta=\omega_{b}$) become asymmetric. In the blue-sideband regime ($\delta=1.3\omega_{b}$), the absorption dip not only transforms into transparency but also exceeds an amplitude of 1, thereby enabling the amplification of the weak probe field. Conversely, in the red-sideband regime ($\delta=0.7\omega_{b}$), there is a lower absorption dip as shown by the blue solid line in~Fig.~\ref{Transmission a}(a). Controlling the magnitude of the magnon driving field can convert MIA into amplification, which has potential uses in quantum sensing and computation.

Notably, the relative phase of the applied fields has been employed to regulate OMIT in cavity optomechanics devices~\cite{PhysRevA.91.043843,PhysRevA.100.013813}. We now study the effect of the relative phase of the cavity and magnon driving fields on the transmission rate of the system. In Fig.~\ref{Transmission a}(b), we set the magnetic field amplitude~$B_{0}=12\times10^{-10}$T and display the transmission spectrum ($|t_{p}|^{2}$) of the output probe field versus the probe detuning with different relative phase ($\Phi$). Figure.~\ref{Transmission a}(b) clearly shows that the absorption resulting from the combined action of magnons and polaritons is sensitive to the relative phase ($\Phi$). Setting~$\Phi=0$ leads to an asymmetric transmission profile in the output spectrum. This configuration results in amplification on the right side of the resonance point ($\delta=\omega_{b}$) and absorption on the left side, as illustrated by the red dotted line in Fig.~\ref{Transmission a}(b). Accordingly, when we set the relative~$\Phi=\pi$ the switching of the asymmetry line shapes becomes noticeable [see blue solid line in~Fig.~\ref{Transmission a}(b)]. In contrast, the central absorption dip at~$\delta=\omega_{b}$ remains fairly stable as the relative phase varies from 0 to~$\pi$. 

The physical mechanism behind this attribute is that by adjusting the relative phase ($\Phi$) of the cavity and magnon driving fields, periodic constructive and destructive interference occurs, significantly altering the cavity-field transmission spectra~\cite{PhysRevA.91.043843,PhysRevApplied.15.024056}. As shown in Fig.~\ref{Transmission a}(b), at the exact resonance point ($\delta=\omega_{b}$), destructive interference occurs, resulting in strong MIA. In contrast, on either side of the resonance point, varying $\Phi$ from 0 to $\pi$ leads to constructive interference and noticeable amplification. This behavior is referred to as magnon-induced amplification (MIAMP). This ability of selectively switching and amplifying
the input probe signal could be highly desirable in practical
optical communications~\cite{PhysRevA.91.043843,Li:17}.

To better illustrate the impact of the relative phase~$\Phi$ on the MMIT profile, we plot the contour map of the transmission spectrum against the optical detuning~($\delta/\omega_{b}$) and the relative phase~$\Phi$ as shown in~Fig.~\ref{Transmission a}(c). The transmission of the probe light changes periodically. At the exact resonance point, we observe strong absorption. The transmission rate transitions from strong absorption to amplification on either side of the resonance point as the phase is tuned from~$-2\pi$ to~$2\pi$. 
  
Therefore, based on the results from Fig.~\ref{Transmission a}, we conclude that the application of an additional magnon driving field in the CMM system allows the MIA and the transparency of the CMM system to be switched by adjusting the magnetic field amplitude and the relative phase while keeping the other system parameters fixed.

To illustrate the significant effect of an OPA on the transmission rate~($|t_{p}|^{2}$) we depict the graph of transmission rate~($|t_{p}|^{2}$) versus detuning ($\delta/\omega_{b}$) as shown in~Fig.~\ref{Angle}. In Fig.~\ref{Angle}(a), in the absence of OPA, specifically when the gain ($G$) along with phase difference ($\theta$) are both zero, we observed that our system displays an absorption dip at resonance position ($\delta=\omega_{b}$) as shown by the red dot curve in Fig.~\ref{Angle}~(a). As anticipated, introducing the nonlinear gain ($G=0.5\kappa_{1}$) results in the MMIT spectrum displaying an asymmetric Fano line shape and an enhanced transparency window. The primary mechanisms generating such Fano-like peaks are the coupling of magnonic modes within cavities and the interference between direct transmission and resonant tunneling paths. Further increasing the nonlinear gain ($G=1\kappa_{1}$) of an OPA improves the transmission rate and wider transparency window, indicating amplification in our system, as indicated by the blue solid curve in Fig.~\ref{Angle}~(a). This is because increasing the nonlinear gain ($G$) of an OPA increases the magnon number, resulting in a broader MMIT window~\cite{PhysRevA.107.063714} and amplification in our system.  In Fig.~\ref{Angle}(b), we plot the transmission rate~($|t_{p}|^{2}$) versus detuning ($\delta/\omega_{b}$) for different phases of an OPA ($\theta = 0.5\pi$, $\theta = \pi$ and~$\theta = 1.5\pi$) we see that phase shift of an OPA also affects the window width as well as the transmission peak. The maximum transmission peak can be obtained when we set the OPA phase $\theta = 0.5\pi$. Hence, we can effectively regulate the transmission rate by modifying the OPA settings.
\begin{figure*}
\centering
\includegraphics[width=0.45\linewidth]{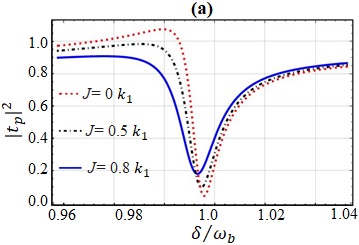}\hspace{2em}
\includegraphics[width=0.45\linewidth]{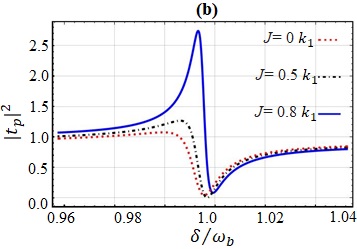}\par\medskip
\includegraphics[width=0.45\linewidth]{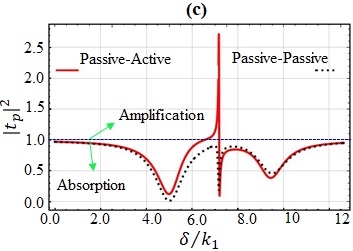}\hspace{2em}
\includegraphics[width=0.45\linewidth]{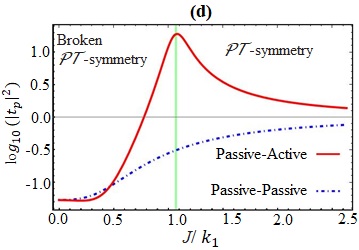}
\caption{~(a$-$b)~The analytical results of the optical transmission rate~($|t_{p}|^{2}$)~versus optical detuning ($\delta/\omega_{b}$) for various optical tunneling strength $J$. (a)~passive-passive~($\kappa_{1}=-\kappa_{2}$)~(b)~passive-active~($\kappa_{1}=\kappa_{2}$).~(c)~Plot of transmission rate~($|t_{p}|^{2}$)~versus optical detuning ($\delta/\kappa_{1}$) with~$J/\kappa_{1}=0.8$. (d)~The logarithm of the transmission rate~($|t_{p}|^{2}$)~versus optical tunneling strength $J/\kappa_{1}$ for passive-passive and passive-active CMM system. With~$J/\kappa_{1}=0.8$ and~$\delta=\omega_{b}$.~All other parameters remains same as in Fig.~\ref{magnon phonon}(c) except for~~$G=0.5\kappa_1$,~$\Phi=0$,~$\theta=\pi/2$  and~$B_{0}=5\times10^{-10}$T.}
\label{active}
\end{figure*}
\subsection{Controllable phase transition and MMIT in CMM system}\label{Controllable phase transition and MMIT in passive-active CMM system}
It is commonly established that systems with $\mathcal{PT}$-symmetry can experience a phase transition, leading to several fascinating and novel phenomena~\cite{PhysRevLett.114.253601,PhysRevA.88.062111}.To investigate the occurrence of a phase transition in our $\mathcal{PT}$-symmetric system, we focus primarily on the optical modes, taking into account both the optical loss and gain. Furthermore, we ignore the influence of OPA, and $g_{1}=g_{2}=g_{mb} =0$. This yields an effective non-Hermitian Hamiltonian, as in~\cite{PhysRevA.105.053705,PhysRevA.108.033517,Wang:24}:

\begin{equation}
\begin{aligned}
H_{eff}= & \hbar \left(\Delta_{a} -{i k_{1}}\right)\hat{a}_{1}^{\dagger} \hat{a}_{1}+ \left(\Delta_{a} +{i k_{2}}\right)\hat{a}_{2}^{\dagger} \hat{a}_{2}\\& + \hbar J \left(\hat{a}_{1}^{\dagger} \hat{a}_{2}+ \hat{a}_{1} \hat{a}_{2}^{\dagger}\right).\label{hamiltonian-222}
\end{aligned}
\end{equation}
Equation~(\ref{hamiltonian-222}) can be
straightforwardly diagonalized via a Bogoliubov transformation~\cite{walls2012quantum}, and we obtain the complex frequency as:
\begin{equation}
\lambda_{ \pm}=\Delta_{a}+\frac{i(\kappa_{1}-\kappa_{2})}{2} \pm \frac{\sqrt{4J^2-\left({\kappa_{1}+\kappa_{2}}\right)^2}}{2}.
\end{equation}
To investigate the $\mathcal{PT}$-broken and $\mathcal{PT}$-symmetric regimes in our system we plot the real and imaginary parts of eigenfrequencies~($\lambda- \Delta_a)/\kappa_1$ against the coupling strength $J/k_{1}$, as depicted in Fig.~\ref{PT}. First, we investigate the equal gain and loss scenario ($k_{1} = k_{2}$), as depicted by the blue-solid and red-dashed curves in Figs.~\ref{PT}(a$-$b). In this scenario, two distinct regimes can be observed: (i) when $J>|\kappa_{1}+\kappa_{2}|/2$ in this scenario the eigenfrequencies of the two 
supermodes exhibit two distinct frequencies ($\Delta_a  \pm \sqrt{4J^2- (k_{1}+k_{2})^2/2}$), and an identical linewidth~($k_{1}-k_{2})/2$, indicating that our system is in a~$\mathcal{PT}$-symmetry regime (as shown by the white region in Figs.~\ref{PT}((a)$-$(b))). (ii) when~$J<|\kappa_{1}+\kappa_{2}|/2$ in this scenario, the
two supermodes coalesce and have the identical frequency and different linewidths, indicating that our system is in a broken~$\mathcal{PT}$-symmetry regime (as illustrated by the light pink shaded area in Figs.~\ref{PT}(a$-$b)).

The critical point, i.e.,~$J=|\kappa_{1}+\kappa_{2}|/2$, where the resonance frequencies and the linewidths of the two supermodes become degenerate, is known as the EP or~$\mathcal{PT}$-phase transition point in our hybrid system. As demonstrated in Figs.~\ref{PT}(a$-$b), by enhancing the coupling rate $J$, one can observe the transition from a broken to an unbroken~$\mathcal{PT}$-phase. Unlike the~$\mathcal{PT}$-symmetric system with gain and loss, if both cavities are operating in a passive-passive mode with the same loss rate, i.e.,~$\kappa_{1}=-\kappa_{2}$, the supermodes always have different frequencies but identical linewidths. The phase transition point does not occur [see black-solid and green-dotted lines in Figs.~\ref{PT}(a$-$b)].

Now, we examine the scenario where the gain/loss are not perfectly matched i.e., $k_{1} = R k_{2}~(R \ne 1)$, where $R$ shows the gain/loss rate. In Figs.~\ref{PT}(c$-$d)  we consider two cases, i.e., $R=0.5$ ($k_{1} =0.5k_{2}$) and $R=2$ ($k_{1} =2k_{2}$). Compared to the equal gain/loss scenario (as in Figs.~\ref{PT}(a$-$b)), one can clearly see that by varying the gain/loss rate ($R=2$ or $R=0.5$) causes the EP to shift the right or left, which implies that either a weaker or a stronger coupling strength $J$ is necessary for $\mathcal{PT}$-phase transition.

The simplification to gain and loss in the cavities is done to isolate and study the fundamental $\mathcal{PT}$-symmetric behavior of the system. This approach allows us to focus on the key aspects of $\mathcal{PT}$-symmetry—gain and loss—without introducing additional complexities that could obscure the desired physical insights. Once we establish a clear understanding
of $\mathcal{PT}$ -symmetry in this context, more complex interactions can be systematically added to
extend the analysis.


In the above subsections~\S\ref{Effect of magnon-photon and magnon-phonon  coupling strength on MMIT window profiles} and~\S\ref{Effect of microwave field and OPA on the transmission and second-order sideband generation}, we examine the impact of various parameters on the transmission rate. However, we keep~$J=0$, which means we work with a single-loss cavity system. Now, our main task is to incorporate the effect of the tunneling rate $J$ and observe its impact on the transmission of our CMM system.

\begin{figure*}
\centering
\includegraphics[width=0.4\linewidth]{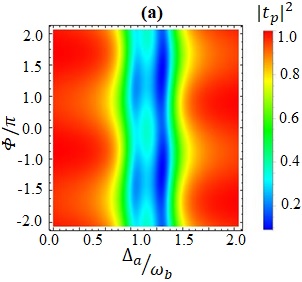}\hspace{3em}
\includegraphics[width=0.4\linewidth]{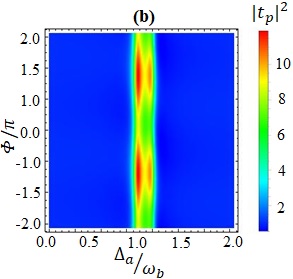}
\caption{~Contour (2D) map of transmission rate~$|t_{p}|^{2}$ vs cavity detuning $\Delta_{a}/\omega_{b}$ and relative phase~$\Phi$.~(a)~passive-passive~($\kappa_{1}=-\kappa_{2}$)~(b)~passive-active~($\kappa_{1}=\kappa_{2}$).~All other parameters remains same as in Fig.~\ref{magnon phonon}(c) except for~$J=1.1\kappa_{1}$,~$\delta=\omega_{b}$,~$G=0.5\kappa_1$,~$\Phi=0$,~$\theta=0$ and~$B_{0}=5\times10^{-10}$T.}
\label{density}
\end{figure*}
Figure~\ref{active}~(a$-$b) displays the plot of the optical transmission rate ($|t_{p}|^{2}$) as a function of optical detuning ($\delta/\omega_{b}$) in the passive-passive and active-passive CMM systems for different optical tunneling strengths $J$, respectively. In~Fig.~\ref{active}(a) for the passive-passive case, we observed that when the optical tunneling strength $J$ is zero, there is a higher absorption dip at the resonance point with a small transmission peak on the left side of the resonance point, as shown by the red dotted curve in Fig.~\ref{active}(a). We noticed that increasing the optical tunneling strength ($J=0.5\kappa_1, 0.8\kappa_1$) for the passive-passive case in our CMM system reduced transmission. This occurs because the increased optical tunneling strength in a passive-passive scenario leads to the formation of two hybrid supermodes,~$\hat{a}_\pm=(\hat{a}_1\pm\hat{a}_2)/\sqrt{2}$, arising from the coupling between the two cavities~\cite{PhysRevLett.114.253601,PhysRevA.101.023842,PhysRevA.95.063825}. The interference between these hybrid supermodes reduces the efficiency of light transmission through the system, thereby compressing the overall transmission.

Compared to the passive-passive scenario (Fig.~\ref{active}(a)), the optical transmission rate in the passive-active CMM system (Fig.~\ref{active}(b)) is significantly boosted. The reason for this enhancement is that as the optical tunneling strength~$J$ increases, the probe light absorption by the passive cavity weakens at resonance, resulting in a transparency window in the spectrum. The increased photon tunneling strengthens the normal-mode splitting between the passive and active cavities, disrupting resonance absorption. Consequently, Stokes scattering is significantly suppressed while the anti-Stokes field (a two-phonon process) is amplified~\cite{Kundu2021}. When the anti-Stokes field becomes degenerates with the probe field, destructive interference between the light pathways in the coupled cavities suppresses absorption, allowing more light to pass through and thus increasing transmission. Therefore, amplification in our CMM system can be tuned by adjusting the optical tunneling strength~$J$~\cite{Kundu2021,PhysRevA.110.023502}.

For better comprehension, we provide the analytical results of the optical transmission rate~$|t_{p}|^{2}$ versus optical detuning~$\delta/\kappa_{1}$ in two types of systems by setting the optical tunneling strength~$J=0.8\kappa_{1}$ as in Fig.~\ref{active}(c). In Fig.~\ref{active}(c), the black-dotted curve illustrates the conventional passive-passive scenario, whereas the red-solid curve represents the passive-active~$\mathcal{PT}$-symmetric-like system. For the passive-passive scenario, an Autler-Townes-splitting-like spectrum is observed~\cite{PhysRevA.98.023821,PhysRevA.102.023707}. However, unlike the passive-passive scenario, a strong signal amplification can be observed between the two Autler-Townes absorption dips in a passive-active scenario. The significant amplification peak is centered at~$\delta=7\kappa_{1}$, as illustrated by the red solid line in Fig.~\ref{active}(c). Figure.~\ref{active}(d) depicts the logarithm of the transmission rate $|t_{p}|^{2}$ vs optical tunneling strength~$J/\kappa_{1}$. For the~$\mathcal{PT}$-symmetric system
with gain and loss (passive-active scenario), as $J$ increases, the signal transmission gradually changes from absorption
to amplification. Significant signal enhancement is achievable at the phase transition point (EP), indicated by the green solid line in Fig.~\ref{active}(d). Moreover, perfect signal absorption can be achieved in the broken $\mathcal{PT}$-symmetric regime. Additionally, when the coupling strength is low, no signal amplification occurs in a system without gain, and significant signal absorption is observed. Consequently, the transmission rate is always less than 1 in the passive-passive system, as shown by the dotted-dashed curve in Fig.~\ref{active}(d). As $J$ increases beyond the phase transition point, the logarithm of the transmission rate forms an approximately straight line near zero. Therefore, in the $\mathcal{PT}$-symmetric CMM system, the absorption and amplification of the probe light can be controlled by adjusting the optical tunneling strength $J$.


Correspondingly, Fig.~\ref{density} shows the contour (2D) map of transmission rate~$|t_{p}|^{2}$ vs cavity detuning $\Delta_{a}/\omega_{b}$ and relative phase~$\Phi$ for two different cases. From Fig.~\ref{density}(a), it is observed that for the passive-passive case, due to the change of relative phase~$\Phi$, a periodic change of the transmission rate takes place, and strong absorption is also observed near the $\Delta_{a}=\omega_{b}$ point. In Fig.~\ref{density}(b), the passive-active scenario shows strong amplification at the $\Delta_{a}=\omega_{b}$ point. The reason for this is that in the passive-active scenario, our $\mathcal{PT}$-symmetric CMM system can add nonlinear effects that increase amplification. These nonlinearities can enhance the strength of the signal in the gain region, resulting in greater overall amplification than systems with solely passive components as in~Fig.\ref{density}(a).   
\begin{figure*}
\centering
\includegraphics[width=0.46\linewidth]{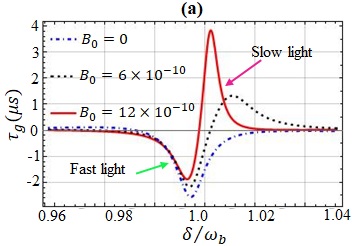}\hspace{2em}
\includegraphics[width=0.46\linewidth]{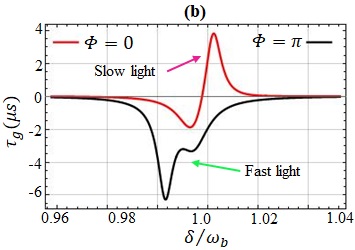}\par\medskip
\caption{Optical group delay~$\tau_g$($\mu s$) of the probe light plotted as a function of optical detuning~$\delta/\omega_{b}$: We set~(a)~The magnetic field amplitude~$B_{0}=0$,~$B_{0}=6\times10^{-10}$T, ~$B_{0}=12\times10^{-10}$T~and relative phase
$\Phi=0$. (b)~The relative phase
$\Phi=0$,~$\Phi=\pi$~and~$B_{0}=12\times 10^{-10}$T~. All other parameters remain same as in Fig.~\ref{magnon phonon}(c). except for $J= 0$,~$\theta=0$,~$G=0$ and~$g_{1}=0$.}
\label{group delay a}
\end{figure*}

\subsection{Optimization of group delays in a compound CMM system}\label{Optimization of group delays in an active-passive compound CMM system}
In general, the CMM systems can not only achieve the MMIT effect but also cause the light to slow or advance~\cite{Kong:19,PhysRevA.107.063714}. This characteristic can be explained through the optical group delay. The optical group delay of the output probe light is~\cite{PhysRevApplied.18.044074,PhysRevA.102.033721,Lu:17}:
\begin{equation}
\tau_g=\frac{d \arg \left(t_p\right)}{d \delta}.
\end{equation}
A group delay greater than zero ($\tau_g>0$) indicates the phenomenon of slow light, whereas a negative group delay ($\tau_g<0$) corresponds to the fast light phenomenon. Specifically, the recent advancements in research utilizing slow light have led to a diverse range of applications, including telecommunications and optical data processing.

The optical group delay ($\tau_g$) of our system's probing field is currently being examined. To illustrate this, in~Fig.~\ref{group delay a} we plot the optical group delay~$\tau_g$($\mu s$) versus optical detuning~$\delta/\omega_{b}$. In Fig.~\ref{group delay a}(a), we depict the group delay~$\tau_g$($\mu s$) of the probe light for differed driving magnetic field amplitudes $B_{0}$ while keeping the relative phase at $\Phi=0$. It is evident that the group delay can be adjusted by varying the magnetic field amplitude of the magnon driving field. When the magnetic field amplitude $B_{0}=0$, a negative group delay is observed [as shown by the blue dashed-dotted line in Fig.~\ref{group delay a}(a)], indicating the fast light effect of the output probe field. Setting the magnetic field amplitude ~$B_{0}=6\times10^{-10}$T reduces the negative group delay at the resonance point and increases the positive group delay on the right side of the resonance point [as shown by the black dotted line in Fig.~\ref{group delay a}(a)], indicating that slow or fast light can be achieved.
As the magnetic field amplitude increases to $B_{0}=12\times10^{-10}$T, the positive group delay rises and reaches a maximum value of $3.8\mu s$ near the resonance point (refer to the red solid line in Fig.~\ref{group delay a}(a)), indicating that the magnon-phonon interaction leads to slow light. Figure.~\ref{group delay a}(b) illustrates that the group delay can be adjusted by modulating the relative phase~$\Phi$ of the applied field. By comparing the group delay profiles for the relative phases~$\Phi=0$, and~$\Phi=\pi$ [as shown by the red and black solid lines in Fig.~\ref{group delay a}(b)], we observe that the positive group delay at the probe detuning~$\delta\approx1.002\omega_{b}$ corresponding to magnon-induced absorption, significantly decreases and shifts in the case of~$\Phi=\pi$. This investigation demonstrates that for a fixed value of the magnetic field amplitude ($B_{0}$), the dispersion property of the system can be modulated, and the slow and the fast light can be achievable and switchable in the CMM system by using the phase modulation of the applied fields.
\begin{figure*}
\centering
\includegraphics[width=0.46\linewidth]{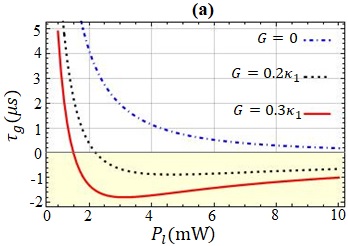}\hspace{2em}
\includegraphics[width=0.46\linewidth]{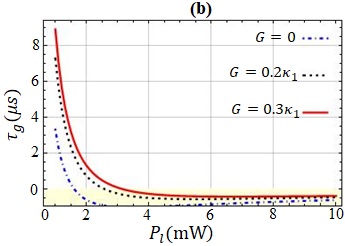}\par\medskip
\caption{Optical group delay~$\tau_g$($\mu s$) of the probe light plotted as a function of pump power~$P_{l}(mW)$, for various values of OPA gains: $G = 0$, $G=0.2 \kappa_1$, and~$G=0.3\kappa_1$. We set~(a)~passive-active system~($\kappa_1 =\kappa_2$)~(b)~passive-passive system~($\kappa_2 = - \kappa_1$).~All other parameters remains same as in Fig.~\ref{magnon phonon}(c). except for $J= 1.06k_{1}$,~$\theta=0$,~$\Phi=0$ and~$B_{0}=5\times10^{-10}$T.}
\label{group delay b}
\end{figure*}

To further investigate the group delay of the probe light in our CMM system, we plot the group delay~$\tau_g$($\mu s$) versus pump power~$P_{l}(mW)$ for various values of OPA gains: $G = 0$~(blue dashed-dot curve), $G=0.2 \kappa_1$~(black dashed curve), and~$G=0.3 \kappa_1$~(red solid curve) as shown in~Fig~\ref{group delay b}. Here, we consider two different scenarios, (i) passive-active system~[as shown in~Fig~\ref{group delay b}(a)],~(ii)~passive-passive system~[as shown in~Fig~\ref{group delay b}(b)].

(i) In~Fig~\ref{group delay b}(a), for~passive-active system~($\kappa_1 =\kappa_2$). In the absence of a degenerate ($G = 0$), the positive group delay of the probe light decreases sharply as the pump power~$P_{l}(mW)$ increases, reaching $\tau_g=0$, and then remains constant with further increases in pump power. This behavior corresponds to the slow light effect [see the blue dashed-dot curve in Fig~\ref{group delay b}(a)]. In sharp contrast, introducing non-linear gain from an OPA ($G=0.2 \kappa_1$, and~$G=0.3 \kappa_1$) into the system and increasing the pump power ($P_{l}(mW)$) causes the group delay ($\tau_g$) of the probe light to become negative  [see the light yellow region in Fig~\ref{group delay b}(a)], resulting in the observation of fast light. One advantage of our CMM system is its ability to gradually produce either fast or slow light by adjusting the OPA gains.

(ii) In~Fig~\ref{group delay b}(b) for~passive-passive system~($\kappa_1 = -\kappa_2$). When compared to the passive-active system, it is evident that the positive group delay has a lower value and decreases sharply as the pump power increases [blue dashed-dot curve in Fig~\ref{group delay b}(b)], indicating that most of the region exhibits negative group delay and the majority of the probe light demonstrates fast-light behavior. Surprisingly, in the presence of an OPA ($G=0.2 \kappa_1$, and~$G=0.3 \kappa_1$), the behavior differs significantly from that of the passive-active system. The positive group delay peak increases with OPA gains, then sharply decreases as the pump power increases up to 
$P_{l}=4(mW)$, after which it remains constant. This suggests that the OPA gains can dramatically affect the dispersion of the system.

In summary, this subsection analyzes the optical group delay of the probe field against probe detuning and pump power, considering different system parameter settings.  Our observations suggest that careful adjustment of system parameters can strongly influence the group delay, thereby modulating the speed of light transmission and enabling a transition from slow to fast light within the system. This tunability offers flexibility in controlling the optical properties and can be leveraged for various applications in optical communication and signal processing~\cite {PhysRevLett.107.133601}. 

\subsection{Possible experimental setup}\label{Possible experimental setup}
In the following subsection, we present numerical values to demonstrate that the assumptions in our analysis are reasonable and consistent with current state-of-the-art experimental achievements, thereby reinforcing the experimental feasibility of our proposal. We consider a microwave cavity containing YIG spheres and an OPA. This method can be experimentally implemented using a planar cross-shaped cavity or a coplanar waveguide made from high-conductivity copper. The choice of cavity structure can affect the frequency range of the magnomechanical interactions studied, as different designs may support various modes and resonances. Planar structures offer benefits in terms of integration with other components, such as magnetic materials or sensors, potentially enhancing the functionality of the device.

The YIG spheres, with a diameter of~$D = 250\mu$m, function as both phonon and magnon resonators, featuring a phonon frequency of~$\omega_b = 2 \pi \times 11.42$~MHz and the spin density~$\rho= 4.22 \times10^{27}\textup{cm}^{-3}$~\cite{doi:10.1126/sciadv.1501286,PhysRevB.105.214418}.  Because of its exceptional material as well as geometrical characteristics, the YIG sphere is also a great mechanical resonator. The phonon along magnon modes is connected to the changing magnetization brought about by the magnon excitation, which deforms the YIG
sphere’s spherical shape as well as vice versa.

To measure group delay experimentally, techniques like interferometry and pump-probe measurements can be employed. Interferometry compares a reference pulse with the output pulse to determine phase shifts related to group delay, while pump-probe methods use a short pump pulse to excite the system and a delayed probe pulse to analyze the response. High-speed photodetectors can capture the temporal characteristics of the output light, enabling precise determination of group delay and insight into the system's dynamics. When comparing time-delay values with state-of-the-art setups, recent experiments have demonstrated highly controllable and tunable slow and fast light effects in various cavity magnonic systems~\cite{PhysRevLett.113.156401,PhysRevApplied.15.024056} achieving group delays on the order of nanoseconds. Notably, the cavity systems in Refs.~\cite{Qin2020,Zhu2024,Lake2020} demonstrate group delays in the microsecond range, which are very close to the values achieved in our system. In our system, by utilizing experimentally realizable parameters, we achieve both pulse advancement and delay, specifically ranging from $\tau_g$ = -7$\mu s$ and $\tau_g$ = 9$\mu s$. This demonstrates excellent agreement with the experimental realizations of fast and slow light reported in Refs.~\cite{Clark2014,Stenner2003} and can be further improved with longer lifetimes in magnonic systems. Recent advancements in cavity magnonics make it feasible to realize the characteristics we have reported in state-of-the-art laboratory experiments, and we believe that the system parameters and structure we propose are achievable with current experimental techniques.



\section{Conclusions}\label{section:Conclusions}
In our work, we have theoretically investigated the features of MMIT, including the transmission rate, and associated group delay, in a two-optically-coupled CMM system. One passive cavity containing an OPA and two YIG spheres to establish magnon-photon coupling. The photon exchange interaction connected the active and passive microwave cavities. We used a perturbation method to establish the analytical formulas for the optical transmission rate. 

Our findings suggest that the transmission spectrum is significantly affected by variations in system parameters. It is worth noting that, unlike the single transparency window observed without magnon-phonon interaction, the presence of magnon-phonon interaction leads to the appearance of two transparency windows and three absorption dips. This phenomenon arises from the combined effects of magnon-phonon and magnon-photon interactions.

Our findings further show that the two absorption dips flanking the central absorption dip may be asymmetrically tuned into amplification and absorption by altering the system parameters.  Notably, through the adjustable coupling strength, by modulating both the loss and gain parameters the system can transition from the $\mathcal{PT}$-symmetry phase to the $\mathcal{PT}$-symmetry-broken phase, which are characterized by distinct normal mode splitting and linewidths. We also demonstrated that the active cavity generates an effective gain, which significantly boosts the transmission of the probe light in the active-passive case compared to the passive-passive case.

Furthermore, by analyzing the group delay of the probe light and adjusting system settings, a transition from slow to fast light can be achieved. We believe that our findings related to the introduced CMM are significant in the sense that these observations can be applied to real-world experiments where controlling transmission light is an important issue to handle, and thus this work may lay the groundwork for future advances in quantum control associated with CMM systems.
\section*{Acknowledgements}\label{section:Acknowledgements}
Abdul Wahab expresses gratitude for the financial support provided by the China Postdoctoral Research Council and Jiangsu University. This work was supported by the Natural Science Foundation of Jiangsu Province (Grant No. BK20231320), National Natural Science Foundation of China (Grant No. 12174157) and Jiangsu Funding Program for Excellent Postdoctoral Talent No.2024ZB867. 
\appendix
\section{Appendix}
\subsection{Calculations of first-order sideband}\label{Appendex}
In this appendix, we present detailed calculations to obtain the amplitudes of the first-order sideband:
\begin{equation}
\begin{aligned}
A_{1+}{h_1}=& i J X_{1+}+i g_{1} M_{1+} +i g_{2} M_{2+} \\&-\sqrt{ \eta_a \kappa_1} \varepsilon_p e^{-i \phi_{pl}}- 2 G e^{i \theta}A_{1-}^{*} ,\\ 
A_{1-}{h_2}=& i J X_{1-}+i g_{1} M_{1-} +i g_{2} M_{2-}- 2 G e^{i \theta}A_{1+}^{*},\\
X_{1+}{h_3}=& i J A_{1+},\\
X_{1-}{h_4}=& i J A_{1-},\\
B_{1+}{h_5}=& i F (M_{2-}^{*}+M_{2+}),\\
B_{1-}{h_6}=& i F (M_{2+}^{*}+M_{2-}),\\
M_{2+}{h_7}=& i g_{2} A_{1+} + i F (B_{1+}+ B_{1-}^{*})- \varepsilon_m e^{-i \phi_{m}},\\
M_{2-}{h_8}=& i g_{2} A_{1-} + i F (B_{1-}+ B_{1+}^{*}),\\
M_{1+}{h_9}=& i g_{1} A_{1+},\\
M_{1-}{h_{10}}=& i g_{1} A_{1-},
\label{Heisenberg-Langevin-7}
\end{aligned}
\end{equation}
 with $h_1 = - i \Delta_{a}+ i \delta- \kappa_1$, $h_2 = - i \Delta_{a}- i \delta- \kappa_1$, $h_3 = - i \Delta_{a}+ i \delta + \kappa_2$, $h_4 = - i \Delta_{a} - i \delta + \kappa_2$,  $h_5 = - i \omega_{b} + i \delta - \kappa_b$, $h_6 = - i \omega_{b} - i \delta - \kappa_b$, $h_7 = - i \Delta_{m2}^{\prime} + i \delta - \kappa_{m2}$, $h_8 = - i \Delta_{m2}^{\prime} - i \delta - \kappa_{m2}$,~$h_9 = - i \Delta_{m1} + i \delta - \kappa_{m1}$, and $h_{10} = - i \Delta_{m1} - i \delta - \kappa_{m1}$.

After solving Eq.~(\ref{Heisenberg-Langevin-7}) for ($A_{1+}$) we get the following constants for first-order side-band:\\
$\alpha_{1} = h_5 h_6^{*} h_7 + F^2 (h_6^{*}-h_5 )$, \\
$\alpha_2 = {h_2}^{*} + \frac{J^2}{{h_4}^{*}} + \frac{g_1^2}{{h_{10}^{*}}}$\\
$\alpha_{3} = h_8^{*} \alpha_{2} + g_{2}^2$,\\
$\alpha_{4} = 2 i G e^{-i \theta} g_{2} + i g_{2} \alpha_{2} $,\\ 
$\alpha_{5} = - i g_{2} \alpha_{4}  - 2 G e^{-i \theta} \alpha_{3}$, \\
$\alpha_{6} =  i g_{2} h_7\alpha_{2}$,\\
$\alpha_{7} = - i g_{2} \alpha_{5} + i g_{2} \alpha_{2} \alpha_{3}$, \\
 $\alpha_{8} =  i g_{2} \alpha_{6} +  \alpha_{2} \alpha_{3} h_7$,\\
$\alpha_{9} = h_8^{*} \alpha_{2} \alpha_{3}$, \\
$\alpha_{10} = \alpha_{1} \alpha_{9}- F^2 (h_6^{*}-h_5) \alpha_{8}$,\\
$\alpha_{11} = i g_{2} h_5 h_6^{*} \alpha_{9} - F^2 (h_6^{*}-h_5) \alpha_{7}$,\\
$\alpha_{12} = \left(\frac{\alpha_{5}\alpha_{10}+ \alpha_{6}\alpha_{11}}{\alpha_{2}\alpha_{3}\alpha_{10}}\right)$, \\ 
$\alpha_{13} = h_1 + \frac{J^2}{h_3} - \frac{i g_1}{h_9} - \frac{i g_2 \, \alpha_{11}}{\alpha_{10}} + 2G e^{-i \theta}  \alpha_{12},
$\\
$\beta_{1} = g_{1}^2 \alpha_{2} - \alpha_{2} \alpha_{3}$, \\
$\beta_{2} = F^2 (h_6^{*}-h_5) \beta_{1} - h_5 h_6^{*} \alpha_{9}$, \\
$\beta_{3} = \left(\frac{\alpha_{6}\beta_{2}}{\alpha_{2}\alpha_{3}\alpha_{10}}- \frac{i g_{1}\alpha_{2}}{\alpha_{2}\alpha_{3}}\right)$, \\ 
$\beta_{4} = \left(\frac{i g_{1}\beta_{2}}{\alpha_{10}}-2G e^{i \theta}\right)$.
{\subsection{Stability analysis}\label{Appendex-1}}
In this appendix we determine the stability of the steady states of our system using the Routh-Hurwitz criterion~\cite{PhysRevA.35.5288}. The fluctuation terms of Eq.~(\ref{Heisenberg-Langevin-3}) in compact matrix form can be rewritten as: 
\begin{equation}
 \delta \dot{\mathbf{x}}=\mathbf{A}\mathbf{x},
\end{equation}
where vector :\\
$\mathbf{x}=(\delta a_1,\delta a_1^{\dagger},\delta m_1, 
\delta m_1^{\dagger},
\delta m_2,
\delta m_2^{\dagger},
\delta b,  
\delta b^{\dagger}, 
\delta a_2 , 
\delta a_2^{\dagger})^T$, in which $T$ denotes the transpose of a matrix. The matrix \( \mathbf{A} \) is
given by:
\begin{widetext}
\[
\mathbf{A} = \begin{pmatrix}
\xi_{1} & 2G e^{i \theta} & -i g_1 & 0 & -i g_2 & 0 & 0 & 0 & -i J & 0 \\
 2G e^{-i \theta} & \xi_{2} & 0 & i g_1 & 0 & i g_2 & 0 & 0 & 0 & i J \\
-i g_1 & 0 & \xi_{3} & 0 & 0 & 0 & 0 & 0 & 0 & 0 \\
0 & i g_1 & 0 & \xi_{4} & 0 & 0 & 0 & 0 & 0 & 0 \\
-i g_2 & 0 & 0 & 0 & \xi_{5} & 0 & -i F & -i F & 0 & 0 \\
0 & i g_2 & 0 & 0 & 0 & \xi_{6} & i F^* & i F^* & 0 & 0 \\
0 & 0 & 0 & 0 & -i F^* & -i F &  -\left(i \omega_b+\kappa_b\right) & 0 & 0 & 0 \\
0 & 0 & 0 & 0 & i F & i F^* & 0 & -\left(-i \omega_b+\kappa_b\right) & 0 & 0 \\
-i J & 0 & 0 & 0 & 0 & 0 & 0 & 0 & -\left(i \Delta_a-\kappa_2\right) & 0 \\
0 & i J & 0 & 0 & 0 & 0 & 0 & 0 & 0 & -\left(-i \Delta_a-\kappa_2\right)
\end{pmatrix}.
\]
\end{widetext}
Where $\xi_{1}=-\left(i \Delta_a+\kappa_1\right)$, $\xi_{2}=-\left(-i \Delta_a+\kappa_1\right)$, $\xi_{3}=-\left(i \Delta_{m_{1}}+\kappa_{m_{1}}\right)$,~$\xi_{4}=-\left(-i \Delta_{m_{1}}+\kappa_{m_{1}}\right)$,~$\xi_{5}= -\left(i \Delta_{m_{2}}^{\prime}+\kappa_{m_{2}}\right)$, and~$\xi_{6}= -\left(-i \Delta_{m_{2}}^{\prime}+\kappa_{m_{2}}\right)$. The characteristic equation $|A-\Lambda I|=0$ can be reduced to $\Lambda^{n}+A_{1}\Lambda^{n-1}+....+A_{n-1}\Lambda + A_{n}=0$ (with $n=10$) where the coefficients can be derived using straightforward but tedious algebra. From the Routh-Hurwitz criterion~\cite{PhysRevA.35.5288},
a solution is stable only if the real part of the corresponding eigenvalue $\Lambda$ is negative~\cite{PhysRevA.100.013813}. These analyses confirm that the experimentally accessible parameters in the main text can keep our compound CMM system in a stable zone.

%
\end{document}